\documentclass[a4paper,5p, sort&compress]{elsarticle}

\usepackage{lineno,hyperref}

\usepackage{subscript}
\usepackage{units}
\usepackage{comment}

\newcommand{\bb}{$0\nu\beta\beta$}
\newcommand{\cd}{$^{113}$Cd }

\begin{document}

\begin{frontmatter}

\title{Long-Term Stability of Underground Operated CZT Detectors Based on the Analysis of Intrinsic $^{113}$Cd $\beta^{-}$--Decay }

  \author[add1]{J. Ebert}
  \author[add2]{C. G\"o\ss ling}
  \author[add3]{D. Gehre\corref{cor1}}
  \ead{daniel.gehre@tu-dresden.de}
  \author[add1]{C. Hagner}
  \author[add1]{N. Heidrich}
  \author[add2]{R. Klingenberg}
  \author[add2]{K. Kr\"oninger}
  \author[add2]{C. Nitsch}
  \author[add1]{C. Oldorf}
  \author[add2]{T. Quante}
  \author[add2]{S. Rajek}
  \author[add1]{H. Rebber}
  \author[add1]{K. Rohatsch}
  \author[add2]{J. Tebr\"ugge}
  \author[add2]{R. Temminghoff}
  \author[add2]{R. Theinert}  
  \author[add1]{J. Timm}
  \author[add1]{B. Wonsak}
  \author[add3]{S. Zatschler}
  \author[add3]{K. Zuber}
  
  \cortext[cor1]{Corresponding author}
  \address[add1]{Universit\"at Hamburg, Institut f\"ur Experimentalphysik, Luruper Chaussee 149, 22761 Hamburg, Germany}
  \address[add2]{Technische Universit\"at Dortmund, Lehrstuhl f\"ur Experimentelle Physik IV, \mbox{Otto-Hahn-Str 4}, 44221 Dortmund, Germany}
  \address[add3]{Technische Universit\"at Dresden, Institut f\"ur Kern- und Teilchenphysik, Zellescher Weg 19, 01069 Dresden, Germany}

\begin{abstract}

The COBRA collaboration operates a demonstrator setup at the underground facility LNGS (Laboratori Nazionali del Gran Sasso, located in Italy) to prove the technological capabilities of this concept for the search for neutrinoless double beta-decay. The setup consists of 64 $(1\times\!1\times\!1)$\,cm$^{3}$ CZT detectors in CPG configuration. One purpose of this demonstrator is to test if reliable long-term operation of CZT-CPG detectors in such a setup is possible. The demonstrator has been operated under ultra low-background conditions since more than three years and collected data corresponding to an exposure of 218\,kg$\cdot$days. The presented study focuses on the long-term stability of CZT detectors by analyzing the intrinsic, fourfold forbidden non-unique $^{113}$Cd single beta-decay. 
It can be shown that CZT detectors can be operated stably for long periods of time and that the $^{113}$Cd single beta-decay can be used as an internal monitor of the detector performance during the runtime of the experiment.

\end{abstract}

\begin{keyword}
\texttt{CZT\sep double beta-decay \sep 113Cd}
\end{keyword}

\end{frontmatter}


\section{Introduction}

Neutrino oscillation experiments have shown that neutrinos have a finite rest mass, but are unable to put a number on how massive they are. The COBRA (Cadmium zink telluride 0-neutrino double Beta-decay Research Apparatus,  \cite{Zuber:2001vm}) collaboration searches for neutrinoless double beta-decay (\bb) to determine the effective Majorana mass of the electron neutrino by a measurement of the  half-life of the $^{116}$Cd decay using CZT (cadmium zinc telluride) semiconductor detectors operated at room temperature. The operation of CZT detectors in a CPG (coplanar grid) configuration \cite{Luke94} enables to use large-volume detectors that contain the isotope of interest and hence results in large intrinsic detection efficiencies. As \bb\,decay is a very rare process with a half-life greater than 1$\times10^{25}$ years in case of $^{116}$Cd, the setup has to be an ultra low-background experiment with a large installed detector mass and which has to be operated for several years to acquire the needed exposure. To reach the required sensitivity, it is necessary to provide extremely stable operation over many years of data taking. One of the main goals of the COBRA demonstrator is to test if a stable long-term operation of CZT-CPG detectors is possible. The stability is tested based on a dataset of 218\,kg$\cdot$days exposure, acquired between Oct'11 and Feb'15.

\section{The COBRA demonstrator}

The COBRA demonstrator \cite{Demo15} consists of 64 CZT-CPG detectors, each with a volume of 1\,cm$^3$ arranged in four layers of 4$\times$4 detectors. The four layers are referred to as L1, L2, L3 and L4 in the following. The total installed detector mass is 380\,grams. 
The continuous operation of the detectors under a controlled, ultra low-background environment allows for an identification of the different background contributions as well as to  develop methods to discriminate such background events from the signal searched for. The detectors are installed in a $(60\times\!60\times\!60)$\,cm$^{3}$ combined low activity lead/copper castle and are kept in a nitrogen atmosphere at room temperature. Furthermore, each detector is permanently powered by its respective bias voltage (BV) and grid bias (GB). 

\section{Measurement strategy}

Since the detectors are made from natural Cadmium containing 12.2\% $^{113}$Cd and as they are operated in an ultra low-background environment, the intrinsic, fourfold forbidden non-unique single beta-decay of $^{113}$Cd$^{\left(\frac{1}{2}+\right)}$$\,\rightarrow\,$$^{113}$In$^{\left(\frac{9}{2}+\right)}$ is by far the strongest signal. The decay of \cd is a direct beta-decay into the ground state of $^{113}$In. 
Hence, the spectrum is continuously spread over the allowed energy range from zero to the \textit{Q}-value of the decay, \textit{Q}\,=\,322.2$\,\pm$\,1.2\,keV \cite{Dawson113}. 
It can be used to monitor the stability of the detectors by analyzing the decay rate of this isotope.
The theoretically expected rate of $^{113}$Cd decays is about 400 per day and detector (assuming 5.9\,g detector mass, 5.0\,at\% zinc concentration and 100\% detection efficiency). Currently, the shape of the $^{113}$Cd spectrum in the range below 100\,keV is not well known and experimental data is strongly requested \cite{Belli113, Mustonen07, Mustonen06}. 
The long half-life of $^{113}$Cd of ($8.00\pm0.35$)$\times10^{15}$ years \cite{Dawson113} and its homogeneous distribution inside the CZT detector should result in a constant decay rate over the time of operation. Changes in the measured decay rate can indicate an alteration of the detector properties which could have an affect on the overall performance of the experiment.

\section{Selection criteria and measurements}
\label{Boundary_conditions}

Due to the uncertainty in the shape of the \cd spectrum, the stability analysis is limited to runs with the lowest common energy threshold for each detector. The energy threshold defines the amplitude of the pulse shape that must be exceeded to trigger the data acquisition system for storing events. The thresholds for the detectors have been adjusted on a run-by-run basis. This is done to reach the lowest possible energy threshold. Thus, the available lifetime depends on the observed energy interval for each detector. Runs with thresholds above 250\,keV are discarded for the stability analysis to avoid too large statistical uncertainties of the detected count rate.
An optimization procedure is applied to identify the optimal threshold \textit{E}$_{\textrm{\begin{small}{opt}\end{small}}}$ to maximize the count rate and to minimize the statistical uncertainties. In this approach, the total number of counts are maximized for each detector as a function of the available lifetime per energy bin. 

\begin{figure} [h]
\begin{center}
 \includegraphics[width=1.0\columnwidth]{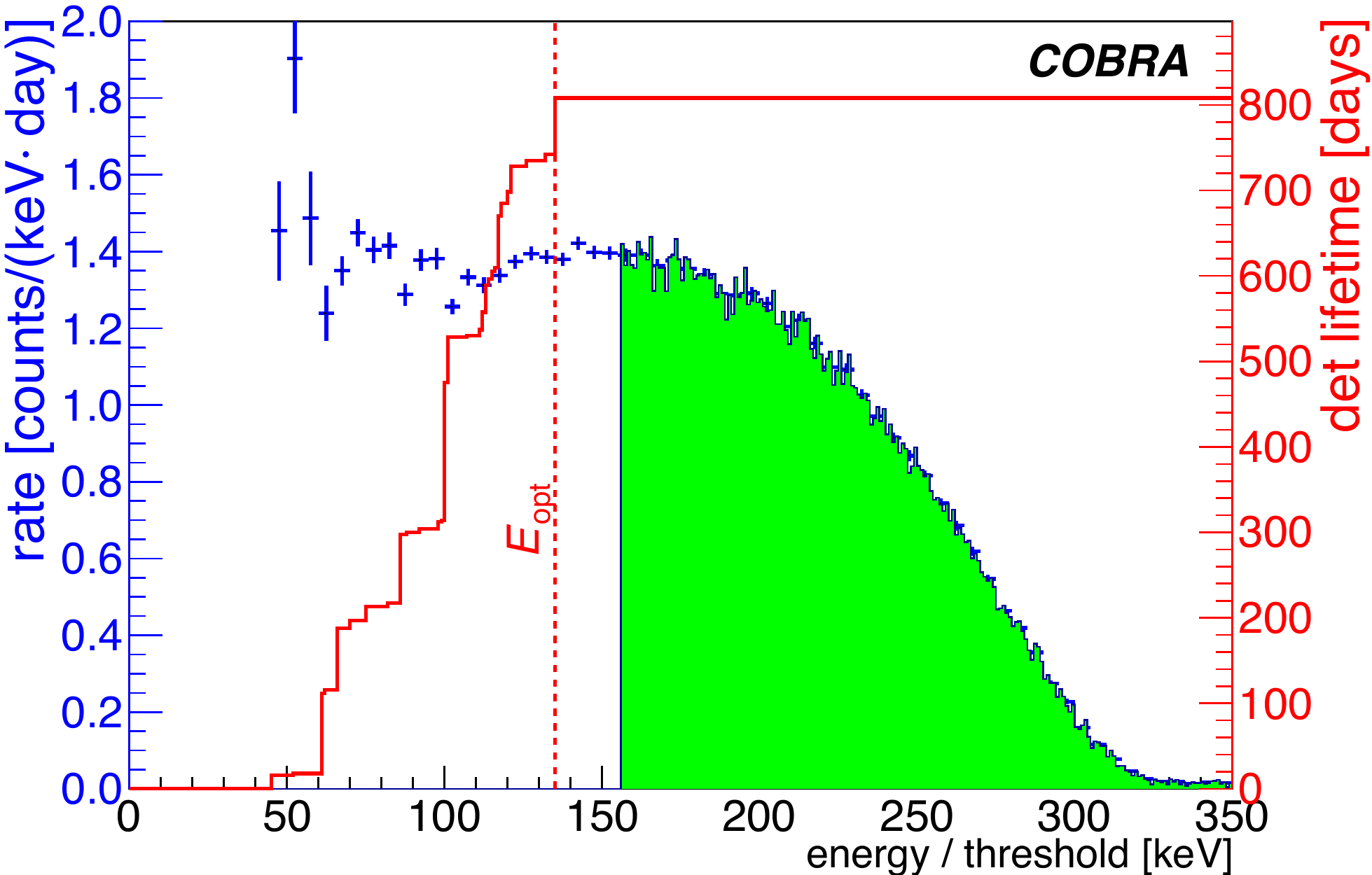}
  \caption{Exemplarily, in blue the total \cd  spectrum of one detector is plotted. The green part of the spectrum is used for the stability analysis. The threshold dependent lifetime of the detector is shown in red. The different lifetimes per energy bins are caused by the operation of the detector at adjusted thresholds in the subsequent runs.}
 \label{L1_total_exposure}
 \end{center}
\end{figure}

In \mbox{Figure \ref{L1_total_exposure}}, the total spectrum (blue) of one detector is shown. The available, threshold-dependent detector lifetime in days is plotted in red. Data taken below the optimal threshold are not used for the stability analysis. The lifetime distribution shows that for lower energies the available lifetime of the detector drops fast and for energies below 60\,keV it is less than 20 days. For the stability analysis, an additional +20\,keV offset is applied to \textit{E}$_{\textrm{\begin{small}{opt}\end{small}}}$. This accounts for the limited energy resolution in that energy range. The remaining part of the spectrum (green histogram) is used for the stability analysis. Runs with energy thresholds higher than \textit{E}$_{\textrm{\begin{small}{opt}\end{small}}}$ are discarded. 

\noindent
A further limitation is caused by the working principle of the coplanar grid. The pulse-shape analysis used in the COBRA demonstrator allows for the reconstruction of the depth of interaction (DOI) as well as to judge if the interaction happened at the lateral surfaces of the detector \cite{Fritts13, Fritts14}. It is known that in the vicinity of the anodes the presence of the grid bias field distorts the movement of the charge carriers. This distortion alters the energy- and depth-reconstruction for energy deposits close to the anodes. For this reason, a second cut (static depth-cut) is used to limit the fiducial volume to the section where the energy and depth are reconstructed correctly. The static depth-cut selects the region $\ge$0.2 and $\le$0.97 of normalized depth and, hence, excludes the effect of the hole shift \cite{Fritts13} in the vicinity of the anodes.

\section{Observed \cd decay rate variations}

The operation under ultra low-background conditions and the analysis of the high-energy part of the spectrum allows to neglect further contributions from other background sources in the analyzed energy range from 100\,keV to 350\,keV for the stability analysis. It is therefore assumed, that the \cd signal region is not contaminated by background. To analyze for the stability of the detected \cd decay rate, the data is partitioned into ten-days real-time intervals. The lifetime of the detectors has to be at least five days for the respective interval to be considered. For each detector, the optimal threshold-cut and the static depth-cut are applied. The remaining \cd counts are summed up and normalized to the lifetime in the respective time period. Due to the different optimal thresholds of each detector, the observed absolute decay-rate deviates from one device to another. 

\begin{figure} [htb]
\begin{center}
  \includegraphics[width=1.0\columnwidth]{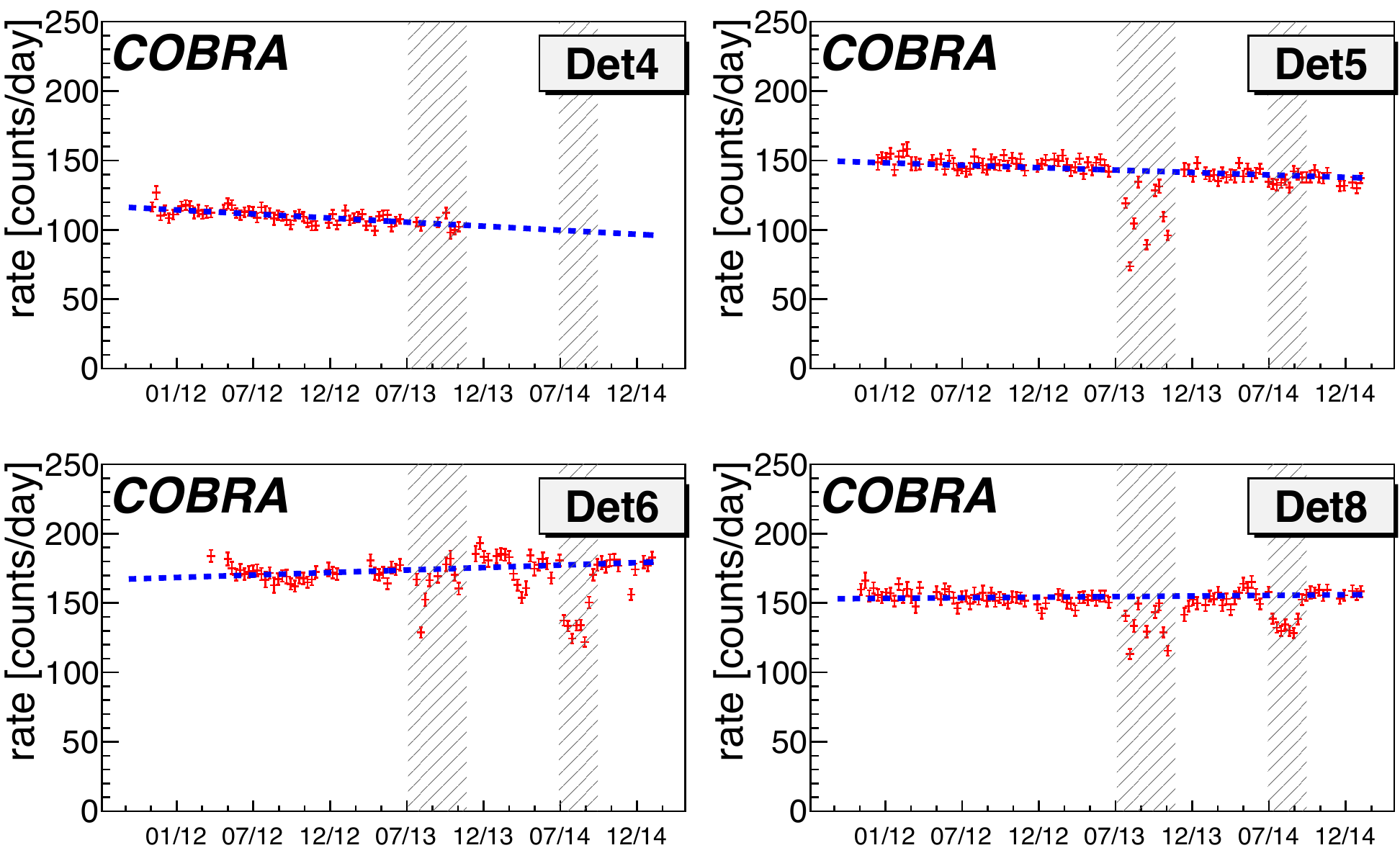}
  \caption{The threshold dependent \cd decay rate of Det4, Det5, Det6 and Det8 of L1. The data is partitioned into ten day intervals, whereas the rate is calculated in counts/day in the analyzed part of the spectrum and varies due to the different optimal thresholds. The uncertainties on the rate are purely statistical in nature. A linear approximation is superimposed in blue. The grey boxes indicate excluded time periods.  }
 \label{decay_rate}
 \end{center}
\end{figure}

The measured \cd decay rates of detectors Det4, Det5, Det6 and Det8 of L1 are shown in Figure \ref{decay_rate} (red marker). The detectors were operational at all times. Missing data points indicate an operation above the optimal energy threshold in the specific time period which led to a removal of the run. A robust, linear approximation is superimposed. The grey boxes indicate excluded time periods (Jul'13 - Nov'13 and Jul'14 - Sep'14), which were affected by electromagnetic interferences that caused an increased baseline noise. This noise led to a higher rejection rate of triggered events and, thus, accidentally removed low-energetic physics events. The \cd rate is especially affected since the deposited energy is always below 322\,keV. Hence, it was decided to exclude the data from these time periods for the evaluation.\

\noindent
Det4 shows the strongest decrease of the $^{113}$Cd rate in L1 ($\Delta_{Det4}$ = --\,5.2\% per year), whereas Det6 is the only detector that shows a slight increase of the count rate \mbox{($\Delta_{Det6}$ = +\,1.4\% per year)}. All other detectors are in between those values for the observed rate variations. For Det8, the measured decay rate of the $^{113}$Cd is basically constant over the full time period (excluding the electronically disturbed runs). This detector yields one of the largest data sets with more than 810 days accepted lifetime. 
As the detectors are produced using material from different batches, the single layers have to be considered independently. The layers were installed subsequently with several months (L1 and L2) or more than one year delay (L3 and L4). Furthermore, L3 is still not performing as well as the other detector layers and has been excluded from the following considerations. This layer is generally more affected by electronic noise and the overall detector performance lags behind those of the other detectors in terms of energy resolution. Additionally, the considered exposure of L3 and L4 is only about one third of those of L1 and L2, which results in fewer time periods and larger uncertainties for the linear approximation. For L1, the average accepted lifetime of the detectors is roughly 700\,days whereas it is for L2 630\,days, for L3 300\,days and for L4 it is only 270\,days.\
\noindent
Nevertheless, the detected count rate variations are relatively low. Beside the excluded time periods no significant outliers are found that would indicate a drastic change of the detector properties or of the data acquisition chain. This can be seen in  \mbox{Table \ref{tab:fit_values}}.

\begin{table} [htb]
\caption{Relative \cd decay rate variations per year for each detector. L1, L2 and L4 are considered for the stability analysis. L3 is excluded due to general performance problems in the low-energy range.}
\begin{small}
\begin{center}
\begin{tabular}{|c|c|c|c|}	

\hline
L1		&	 Rel.rate change/yr 	&	L2		&	Rel.rate change/yr \\
\hline	
Det1		&	1.000$\pm$0.004	&	Det17	&	1.001$\pm$0.001\\
Det2		&	0.977$\pm$0.016	&	Det18	&	1.010$\pm$0.010\\
Det3		&	0.960$\pm$0.019	&	Det19	&	0.993$\pm$0.001\\
Det4		&	0.948$\pm$0.052	&	Det20	&	1.006$\pm$0.006\\
Det5		&	0.976$\pm$0.024	&	Det21	&	1.025$\pm$0.025\\
Det6		&	1.014$\pm$0.015	&	Det22	&	0.973$\pm$0.016\\
Det7		&	0.996$\pm$0.004	&	Det23	&	0.967$\pm$0.018\\
Det8		&	1.000$\pm$0.002	&	Det24	&	1.015$\pm$0.015\\
Det9		&	1.005$\pm$0.005	&	Det25	&	0.991$\pm$0.012\\
Det10	&	0.998$\pm$0.005	&	Det26	&	1.035$\pm$0.032\\
Det11	&	0.984$\pm$0.016	&	Det27	&	1.003$\pm$0.007\\
Det12	&	1.004$\pm$0.004	&	Det28	&	1.033$\pm$0.033\\
Det13	&	0.999$\pm$0.001	&	Det29	&	1.031$\pm$0.029\\
Det14	&	1.001$\pm$0.003	&	Det30	&	0.969$\pm$0.031\\
Det15	&	0.998$\pm$0.002	&	Det31	&	0.961$\pm$0.022\\
Det16	&	0.943$\pm$0.057	&	Det32	&	1.056$\pm$0.056\\
\hline
L3		&	 Rel.rate change/yr 	&	L4		&	Rel.rate change/yr \\
\hline					
Det33	&	0.982$\pm$0.001	&	Det49	&	0.980$\pm$0.020\\
Det34	&	1.004$\pm$0.014	&	Det50	&	0.997$\pm$0.012\\
Det35	&	0.966$\pm$0.034	&	Det51	&	0.980$\pm$0.020\\
Det36	&	0.695$\pm$0.305	&	Det52	&	1.030$\pm$0.030\\
Det37	&	0.976$\pm$0.024	&	Det53	&	1.032$\pm$0.032\\
Det38	&	0.891$\pm$0.109	&	Det54	&	0.985$\pm$0.006\\
Det39	&	0.892$\pm$0.108	&	Det55	&	0.951$\pm$0.006\\
Det40	&	0.928$\pm$0.072	&	Det56	&	1.042$\pm$0.031\\
Det41	&	0.960$\pm$0.010	&	Det57	&	0.951$\pm$0.049\\
Det42	&	0.849$\pm$0.070	&	Det58	&	0.983$\pm$0.002\\
Det43	&	0.916$\pm$0.035	&	Det59	&	0.908$\pm$0.032\\
Det44	&	0.964$\pm$0.004	&	Det60	&	1.022$\pm$0.022\\
Det45	&	0.839$\pm$0.161	&	Det61	&	0.978$\pm$0.002\\
Det46	&	0.857$\pm$0.071	&	Det62	&	0.989$\pm$0.011\\
Det47	&	0.705$\pm$0.295	&	Det63	&	1.012$\pm$0.012\\
Det48	&	0.733$\pm$0.267	&	Det64	&	1.038$\pm$0.038\\
\hline

\end{tabular}
\end{center}
\end{small}

\label{tab:fit_values}													
\end{table}

\begin{figure} [htb]
\begin{center}
 \includegraphics[width=1.0\columnwidth]{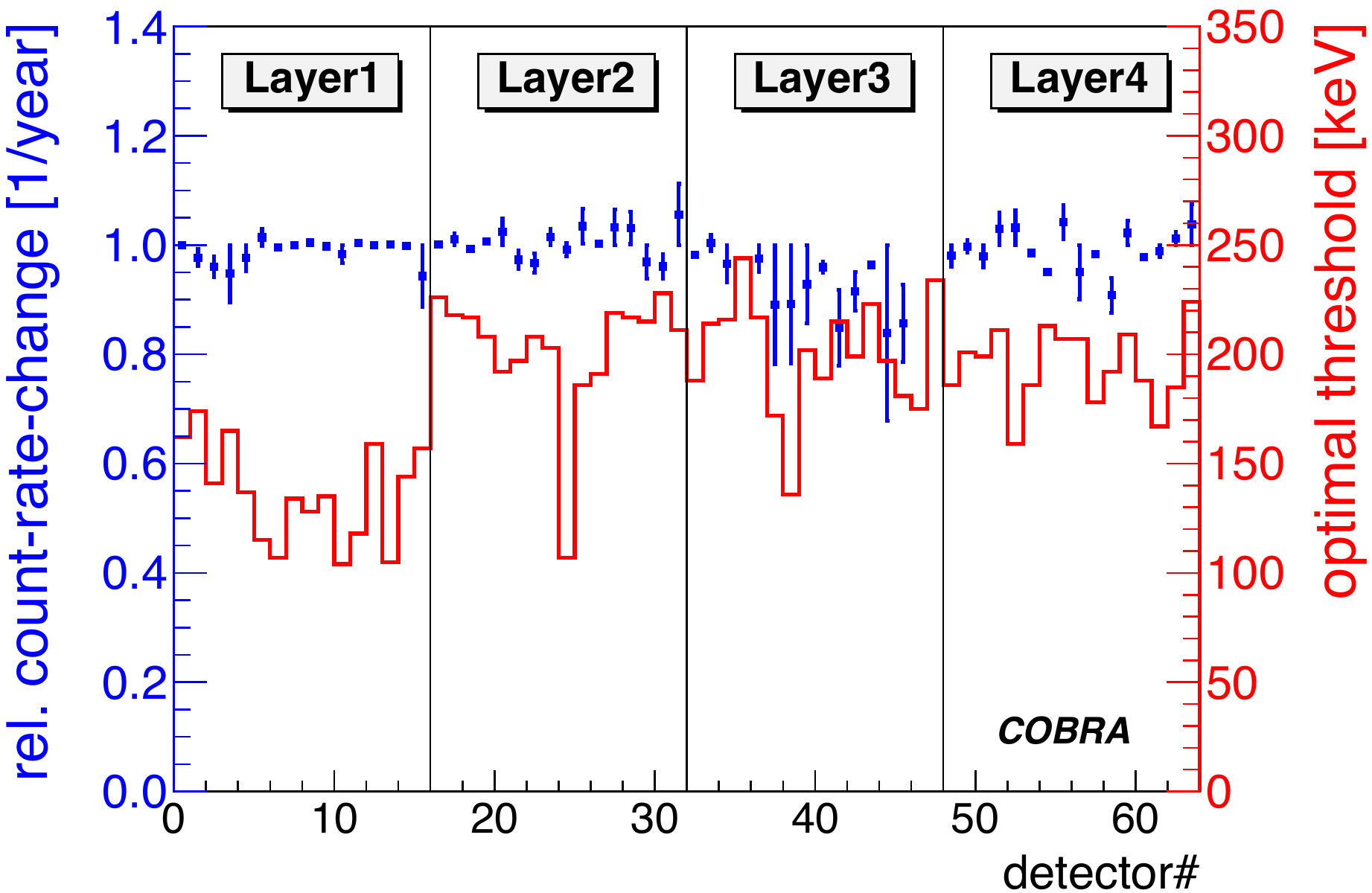}
  \caption{The relative \cd count rate change on a per year basis is plotted in blue. In red the respective optimal threshold per detector is shown. }
 \label{count_rate_eth}
 \end{center}
 \vspace{3mm}
\end{figure}

The relative change in count rate per year and detector is plotted in \mbox{Figure \ref{count_rate_eth}}. The error bars indicate the uncertainty of the linear approximation on the data which is basically related to statistical uncertainties. 45 out of 48 detectors from L1, L2 and L4 show a change in the decay rate well below $\pm$6\% per year. The shorter overall lifetime and the higher optimal thresholds of the detectors of L4 affects the accuracy of the approximation as only a smaller part of the spectrum is analyzed and fewer accepted run periods are available. Nevertheless, a weighted mean of $\bar{p}_1=(0.995\pm$0.004) can be reported, which indicates an almost perfect stable operation of the detectors.

\begin{figure} [htb]
\begin{center}
 \includegraphics[width=1.0\columnwidth]{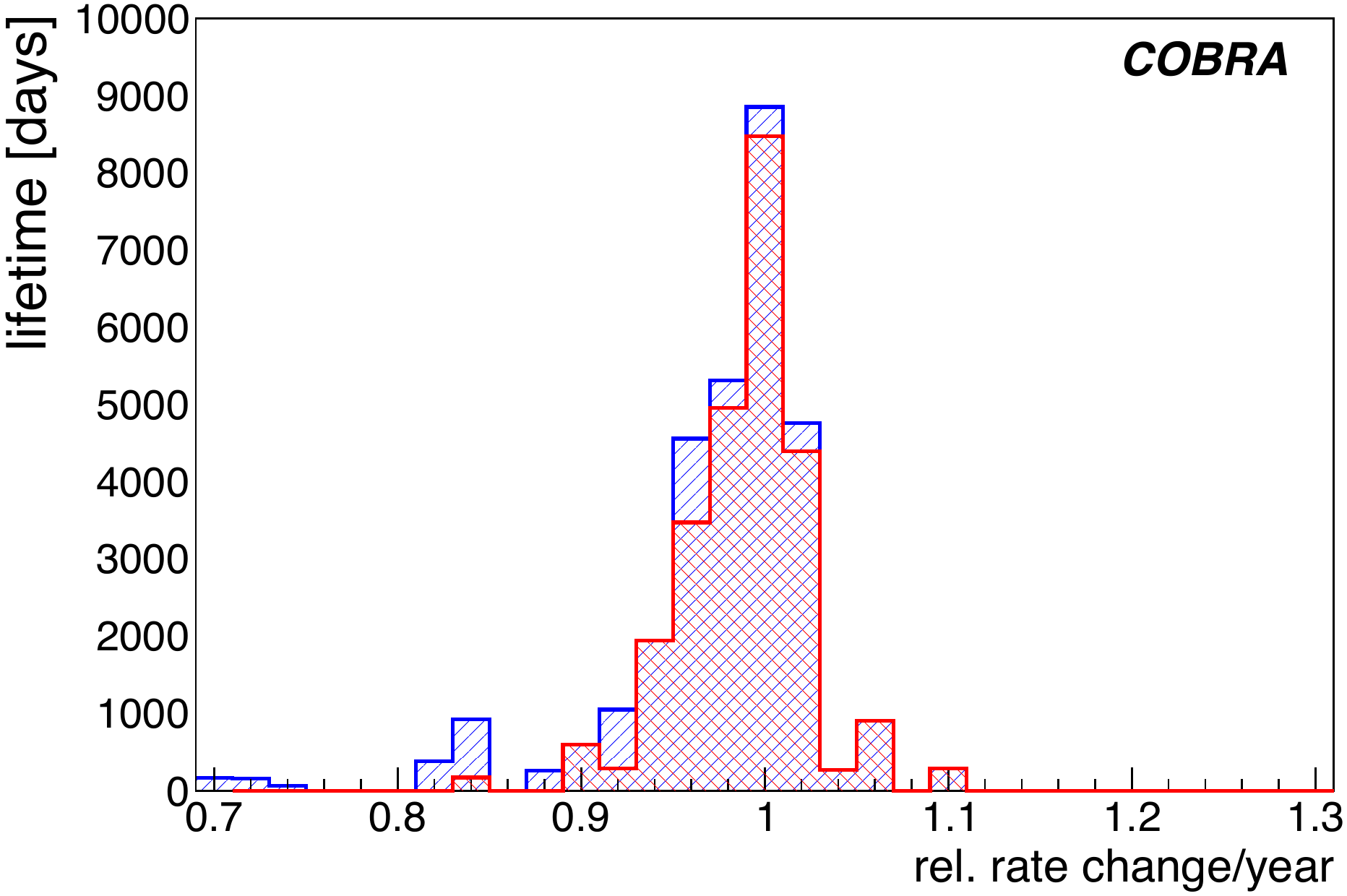}
  \caption{The lifetime-weighted distribution of the relative rate change of all Layers (blue) and Layer1, Layer2 and Layer4 in red. }
 \label{rate_change_histo}
 \end{center}
\end{figure}

Figure \ref{rate_change_histo} displays the lifetime-weighted relative change rate distribution for all layers (blue) and for L1, L2 and L4 in red. 
The distribution is clustering around one with a slight shift towards values less than one.

\begin{figure} [htb]
\begin{center}
 \includegraphics[width=1.0\columnwidth]{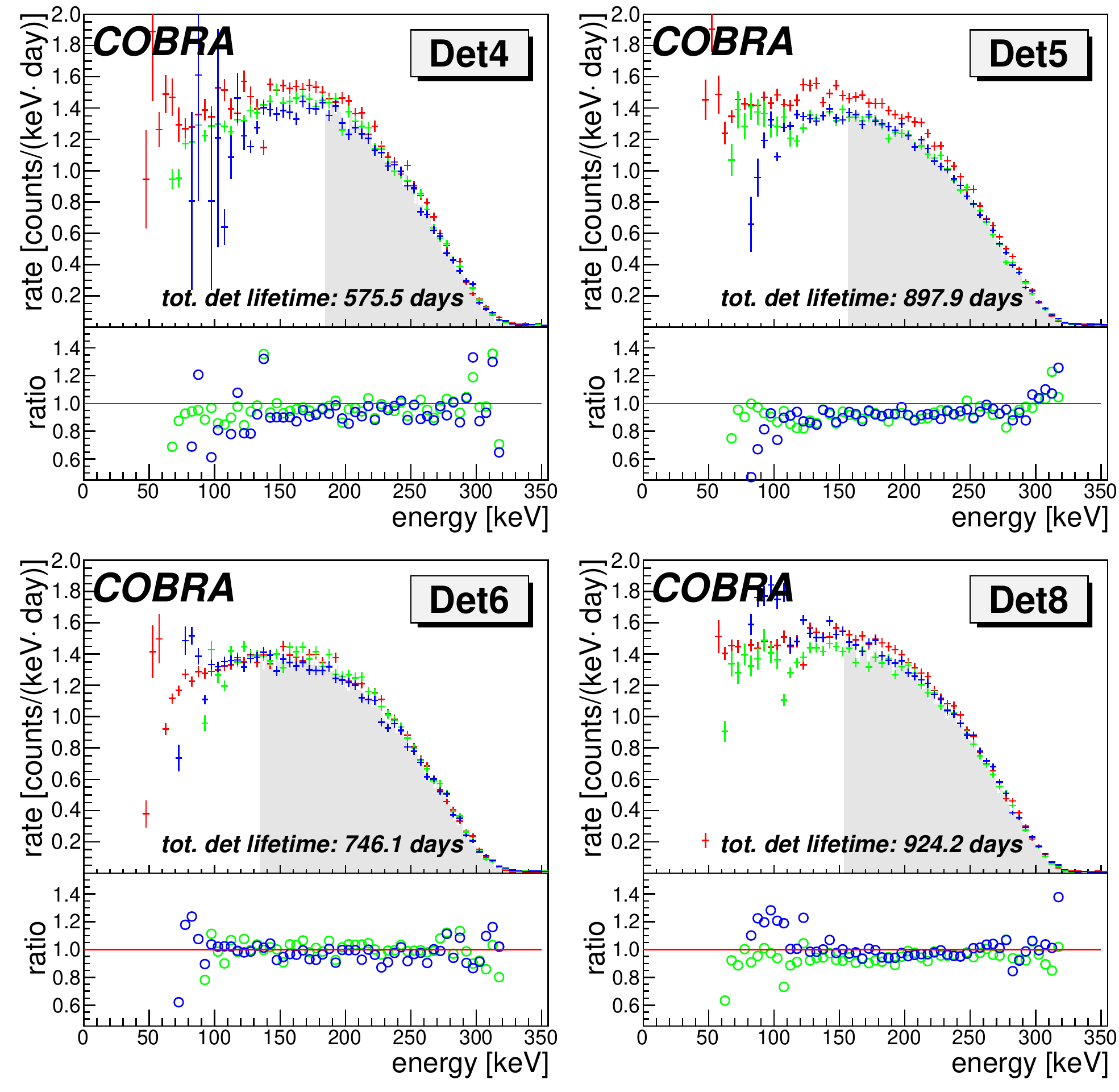}
   \caption{Comparison of the spectral shapes of the \cd decay for different time sections. The data is partitioned into three time periods: first third (red), second third (green) and last third (blue). The grey part of the spectrum has been used for the stability analysis. The lower inset shows the relative deviation of the middle (green) and the end period (blue) with respect to the start period (red).}
 \label{spectral_shape}
 \end{center}
 \vspace{-3mm}
\end{figure}

To analyze for the cause of the observed changes, the spectral shape and the depth distribution of different time periods of the experiment are compared. To accomplish this, the number of available intervals is sectioned into three parts: the start, middle and end of the lifetime. For the three time periods, the counts per keV are summed up bin-wise and are normalized to the respective lifetime per bin for each detector. In Figure \ref{spectral_shape}, the three spectra of the detectors are plotted in red (first third), green (second third) and blue (last third). In grey, the part of the spectrum that lies above the (\textit{E}$_{\textrm{\begin{small}{opt}\end{small}}}$+20)\,keV is marked. Only this part is used for the stability analysis. 
In the presented cases the red spectra (first third of the data taking period) tends to lie slightly above the green and blue ones - but no clear tendency is observable if one takes the other detectors into account. In general, no shift or deformation of the spectral shape has been found comparing the different time periods. 

\begin{figure} [tb]
\begin{center}
 \includegraphics[width=1.0\columnwidth]{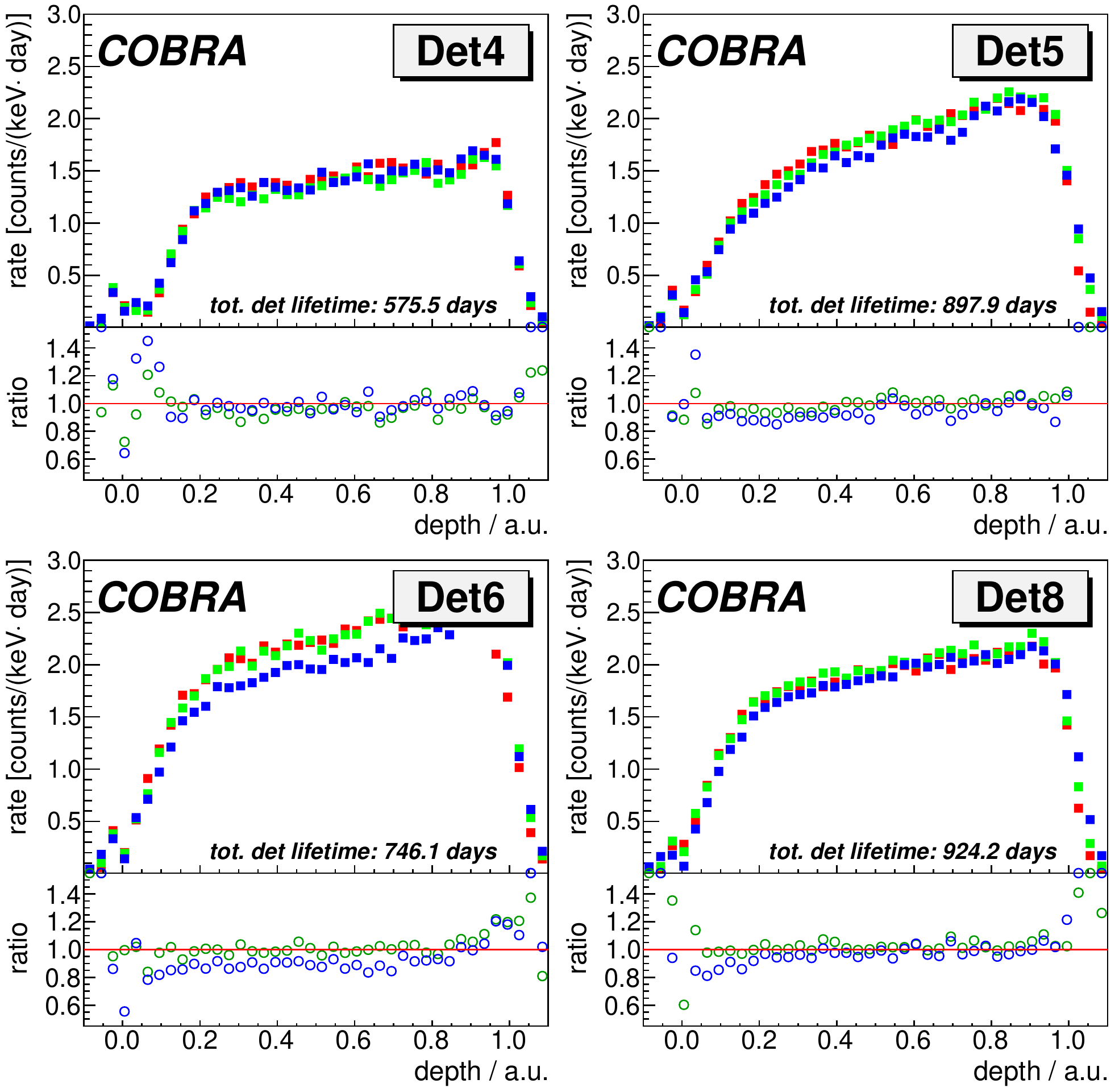}
  \caption{Comparison of the depth distribution of the \cd decay for different time sections (anodes at '0', cathode at '1'). For the stability analysis the static depth cut has been applied.}
 \label{depth_distribution}
 \end{center}
  \vspace{-3mm}
\end{figure}

For the stability analysis of the reconstructed depth distribution, the same time sectioning and least common energy threshold-cuts are applied. In contrast to the decay-rate based stability analysis that implies the static depth-cut, no depth dependent cuts are applied here. This allows for an analysis of the full depth distribution of the recorded interactions. No time-correlated impact on the reconstructed depth is visible as can be seen in \mbox{Figure \ref{depth_distribution}}. The fact that the shape of the depth distribution looks different for the single detectors is caused by their different thresholds. The lower the threshold, the higher the total count rate. The asymmetry of the reconstructed depth, which shows a shift of the reconstructed events towards the cathode, is caused by imprecisions of the reconstruction algorithm for lower energies.

\section{Conclusion}

The presented study is based on the analysis of the intrinsic, fourfold forbidden non-unique single beta-decay of \cd with a \textit{Q}-value of \textit{Q}=(322.2$\,\pm$\,1.2)\,keV. The result obtained is limited to the energy range from approximately 100\,keV to 350\,keV. The analyzed dataset represents a lifetime of roughly 3.5 years with a total acquired exposure of 218\,kg$\cdot$days, which is the longest measurement ever taken with CZT detectors under these conditions. \\
\noindent
Based on this dataset, count rate variations in the lower percent-per-year range ($\Delta_{mean}=(-0.5\pm0.4)$\,\% per year) are found for 45 out of 48 detectors. The majority of the detectors are working without any glitches or sudden performance changes. The reproduction of the spectral information as well as the depth reconstruction of the detectors is not affected and is stable over the analyzed time period. The 16 detectors of layer L3 are excluded due to known performance issues. Disregarding L3, the overall performance and stability of the setup is excellent. Currently it is not clear what causes the observed minor alterations of the count rate and if this is seen for higher energies as well. To analyze for efficiency alterations in higher energy ranges, a continuous monitoring of the detectors based on precisely positioned calibration sources is required.
Nevertheless, under the precondition of ultra low-background operation, the analysis of the intrinsic \cd decay offers a unique way to monitor the detector performance during the runtime of the experiment.\\
\noindent
Finally, it can be concluded that CZT detectors can be operated stably under ultra low-background conditions over time scales of at least several years. This result shows that it is possible to move towards a large-scale experiment as proposed by the COBRA collaboration. 

\vspace{9cm}

\section{Acknowledgements}

The authors would like to thank the funding agency ``Deutsche Forschungsgemeinschaft" (DFG Zu 123/3-4 \& \mbox{Zu 123/15-1}) that supports the development and operation of the COBRA demonstrator and the Laboratori Nazionali del Gran Sasso for hosting the experiment.

\section*{References}
\bibliography{113cd_stability}

\end{document}